\documentclass[manuscript, review, screen]{article}
\usepackage{amssymb}
\usepackage{algorithm}
\usepackage{algpseudocode}
\usepackage{ntheorem}
\usepackage{graphicx}

\usepackage[table]{xcolor}
\usepackage{yhmath}

\providecommand{\keywords}[1]
{
	\small	
	\textbf{\textit{Keywords---}} #1
}

\begin{document}

	\title{SARRIGUREN: a polynomial-time complete algorithm for random $k$-SAT with relatively dense clauses}
	\author{Alfredo Go\~{n}i Sarriguren \\
		Department of Computer Languages and Systems \\
		Faculty of Informatics \\
		University of the Basque Country UPV/EHU\\
		Paseo Manuel de Lardizabal 1 (20018) Donostia, Spain\\
		alfredo@ehu.es}
	
\maketitle
	
	\begin{abstract}
	
SARRIGUREN, a new complete algorithm for SAT  based on counting clauses (which is valid also for Unique-SAT and \#SAT) is described, analyzed and tested. Although existing complete algorithms for SAT perform slower with clauses with many literals, that is an advantage for SARRIGUREN, because the more literals are in the clauses the bigger is the probability of overlapping among clauses, a property that makes the clause counting process more efficient.
Actually, it provides a
$O(m^2 \times n/k)$  time complexity for random $k$-SAT instances of $n$ variables and $m$ relatively dense clauses, where that density level is relative to the number of variables $n$, that is, clauses are relatively dense when $k\geq7\sqrt{n}$. 
Although theoretically there could be worst-cases with exponential complexity, the probability of those cases to happen in random $k$-SAT with relatively dense clauses is practically zero.
The algorithm has been empirically tested and that polynomial time complexity maintains also for $k$-SAT instances with less dense clauses ($k\geq5\sqrt{n}$).
That density could, for example, be of only 0.049 working with $n=20000$ variables and $k=989$ literals. 
In addition, they are presented two more complementary algorithms that provide the solutions to $k$-SAT instances and valuable information about number of solutions for each literal.
Although this algorithm does not solve the NP=P problem (it is not a polynomial algorithm for 3-SAT), it broads the knowledge about that subject,
because $k$-SAT with $k>3$ and dense clauses is not harder than 3-SAT.
Moreover, the Python implementation of the algorithms, and all the input datasets and obtained results in the experiments are made available.
\end{abstract}


\section{Introduction}
\label{Intro}

Boolean satisfiability problem (SAT) stands as an iconic challenge that has been extensively researched \cite{Biere2009}, since it was found to be the first NP-complete algorithm \cite{cook1971}. 
A reduced version of SAT is $k$-SAT where all clauses are represented in CNF (Conjunctive Normal Form) and contain exactly $k$ literals. Although  algorithms to run in polynomial-time have been defined for 2-SAT \cite{Krom1967TheDP,doi:10.1137/0205048}, $k$-SAT remains to be NP-complete for $k\geq 3$. The complexity of current complete algorithms that solve $k$-SAT is exponential: in \cite{DANTSIN200269} it is 
described a deterministic local search algorithm for $k$-SAT that runs in  $O((2-\frac{2}{k+1})^n)$ time, what means that it runs in $O(1.5^n), O(1.6^n),$ and $O(1.\wideparen{6}^n)$ time for values of $k$ equal to 3, 4, and 5 respectively. In \cite{10.1016/j.tcs.2004.08.002} a faster but still exponential $O(1.473^n)$ algorithm for 3-SAT is presented. 
Moreover, in \cite{10.1145/3313276.3316359} they present an algorithm $O(1.307^n)$ for Unique 3-SAT (a variation of 3-SAT that decides if there is only one satisfying assignment or not). It seems that the complexity 
of $k$-SAT grows for higher values of $k$, what sounds reasonable because known complete algorithms (Davis-Putnam' \cite{10.1145/321033.321034}, Stalmarck’s algorithm \cite{STALMARCK199031}, DPLL \cite{10.1145/368273.368557} and others based on them) make use of existential quantifications, inference rules such as resolution and search that perform slower with higher values of $k$, that is, with more literals in the clauses. In fact, in \cite{IMPAGLIAZZO2001367} authors claim that the complexity of $k$-SAT increases with increasing $k$, but this is only under the assumption that $k$-SAT does not have subexponential algorithms for $k \geq 3$; an assumption that has to be revisited, according to this work.

In this paper a new complete algorithm for SAT named SARRIGUREN\footnote{SARRIGUREN is a Basque name of a town in Nafarroa/Navarra  whose etymological meaning is ``beautiful thicket". Moreover, it is also my mother's family name.} is presented. That algorithm does not make use of existential quantifications, inference rules nor search. It is based on counting unsatisfiable variations for clauses, and the fact of having many literals in the clauses (big values of $k$)
results in an advantage that will be shown in this paper. That algorithm provides an $O(m^2 \times \frac{n}{k})$ polynomial-time complexity when applied to
random $k$-SAT with dense clauses\footnote{As it will be discussed in section \ref{worstCase}, the probability of exponential time complexity for some worst-cases in random $k$-SAT with dense clauses is practically zero.}, that is, to a set of $m$ clauses of exactly $k$ randomly-chosen literals of $n$ different variables where $k$ is relatively close to $n$ ($k\geq 7\sqrt{n}$ or even 
$k\geq 5\sqrt{n}$). Moreover, this complete algorithm based on counting unsatisfiable variations of clauses is also a complete algorithm for variations of SAT explained in \cite{Biere2009} such as the previously mentionated Unique-SAT or propositional model counting \#SAT (a variation of SAT that calculates the total number of satisfying assignments).

There are also many incomplete methods that are very efficient
in finding solutions to many instances of SAT, but that cannot
guarantee unsatisfiability even if they do not find a solution. Among them we can find Survey Propagation that can solve
random 3-SAT instances with one million variables and beyond
in near-linear time \cite{doi:10.1126/science.1073287}. Up to my knowledge no performance results have been reported for $k$-SAT instances with higher values of $k$.

In the following sections the algorithm is explained with an example, some definitions and concepts related with the algorithm are given and proven, the algorithm is presented, its corresponding analysis of complexity, experimental results of the algorithm, another algorithm to get the solutions or satisfying assignments is also explained, and finally, the conclusions are presented.

\section{Explanation of the algorithm with an example}
	
Let us consider two sets of clauses: $K_1=\{k_1,k_2,k_3,k_4\}$ that is satisfiable and $K_2=\{k_1,k_2,k_3,k_4,k_5,k_6\}$ that is unsatisfiable, where the clauses are:

\begin{center}
$k_{1} = \overline{x_{1}} \lor {x_{2}}$

$k_{2} = {x_{2}} \lor {x_{3}}$

$k_{3} = \overline{x_{2}} \lor \overline{x_{3}}$

$k_{4} = {x_{1}} \lor \overline{x_{2}} \lor {x_{3}}$

$k_{5} = {x_{1}} \lor {x_{2}}$ 

$k_{6} = \overline{x_{1}} \lor \overline{x_{2}}$ 
\end{center}

It is known that $K_1$ is satisfiable if there is at least a variation\footnote{This term variation refers a variation with repetition of 2 Boolean values taken $n$ variables at a time.} of Boolean values for the variables (${x_{1}}, {x_{2}}, {x_{3}}$) such as $K_1 = k_{1} \land  k_{2} \land  k_{3} \land  k_{4} = 1$. In Table~\ref{truthTable} it can be seen that there are two variations (the satisfying assignments or solutions) that make the set of clauses $K_1$ satisfiable: $<0,0,1>$ and $<1,1,0>$. When the clauses $k_{5}$ and  $k_{6}$ are added, then $K_2$ is unsatisfiable because the combinations $<0,0,1>$ and $<1,1,0>$ make $k_{5}$ and $k_{6}$ equal to 0, respectively.

\begin{table}[h]
	\centering
	\begin{tabular}{|c|c|c||c|c|c|c||c||c|c|c||c|}		
		\hline   $x_{1}$ & $x_{2}$  &  $x_{3}$ & $k_{1}$ & $k_{2}$ & $k_{3}$ & $k_{4}$ & $K_{1}$
		 & $\cdots$ & $k_{5}$ & $k_{6}$ & $K_{2}$ \\
		\hline   0 & 0  & 0 & 1 & 0 & 1 & 1& 0 & $\cdots$ &  0 & 1 & 0\\		
		\hline  \cellcolor{blue!25}0 &  \cellcolor{blue!25}0  &  \cellcolor{blue!25}1 & 1 & 1& 1 & 1& \cellcolor{blue!25}1 & $\cdots$ & \cellcolor{blue!25}0 & 1 & \cellcolor{blue!25}0\\		
		\hline   0 & 1  & 0 & 1 & 1 & 1 & 0& 0 & $\cdots$ & 1 & 1 & 0\\		
		\hline   0 & 1  & 1 & 1 & 1 & 0 & 1& 0 & $\cdots$ & 1 & 1 & 0\\		
		\hline   1 & 0  & 0 & 0 &  0 & 1 & 1& 0 & $\cdots$ & 1 & 1 & 0\\		
		\hline   1 & 0  & 1 & 0 &  1 & 1 & 1& 0 & $\cdots$ & 1 & 1 & 0\\		
		\hline   \cellcolor{blue!25}1 &  \cellcolor{blue!25}1  &  \cellcolor{blue!25}0 & 1 & 1 & 1 & 1& \cellcolor{blue!25}1 & $\cdots$ &  1 &\cellcolor{blue!25}0 & \cellcolor{blue!25}0\\		
		\hline   1 & 1  & 1 & 1 & 1 & 0 & 1& 0 & $\cdots$ & 1 & 0 & 0\\	
		\hline	
	\end{tabular} 
	\caption{Truth table for $K_1$ and $K_{2}$}
	\label{truthTable}
\end{table}

It is not an efficient algorithm for solving SAT to build all the variations and check if each variation makes false at least one of the clauses because it has a complexity of $O(2^{n})$, where $n$ is the number of variables. However, to {\it count} the number of variations that make unsatisfiable at least one clause is much more efficient for many instances of SAT.
If $u$ is the number of unsatisfying variations and is equal to $2^{n}$, then the set of clauses is unsatisfiable. And, in other case, it is satisfiable and there are exactly $(2^{n}-u)$ solutions. Following with the previous example, notice that 
$K_{2}$ is unsatisfiable because the $2^{3}=8$ variations do not satisfy at least one clause, and $K_1$ is satisfiable because only 6 unsatisfiable variations have been found, what also means that there are $(2^{3}-6)=2$ solutions that satisfy all clauses.
 
Let us start by analyzing how can be counted the unsatisfying variations for each clause:

\begin{itemize}
\item $k_{1} = \overline{x_{1}} \lor {x_{2}}$ $\Rightarrow$ the set $k_{1}^{full}=\{ \overline{x_{1}} \lor {x_{2}} \lor \overline{x_{3}},
\overline{x_{1}} \lor {x_{2}} \lor x_{3}\}$ is equivalent to $k_{1}$ $\Rightarrow$ $\{<1,0,1>,<1,0,0>\}$ or
$\{ x_{1} \land \overline{x_{2}} \land x_{3},
x_{1} \land \overline{x_{2}} \land \overline{x_{3}}\}$
is the set of unsatisfying variations or complementary conjunctive clauses that unsatisfy the clauses in $k_{1}^{full}$. That set of variations follows the 
pattern $P_{1}=<1,0,->$. If  $v$ is the number of `$-$' in the pattern or the number of variables that do not appear in the clause  $k_{1}$, then there are exactly $2^{v}=2^1=2$ unsatisfying variations; and exactly the same number of clauses in $k_{1}^{full}$. 
Moreover, the complementaries of $(k_{1}^{full})^C$ 
$=\{ 
x_{1} \lor {x_{2}} \lor x_{3},
x_{1} \lor {x_{2}} \lor \overline{x_{3}},
x_{1} \lor \overline{x_{2}} \lor x_{3},
x_{1} \lor \overline{x_{2}} \lor \overline{x_{3}},
\overline{x_{1}} \lor \overline{x_{2}} \lor x_{3},
\overline{x_{1}} \lor \overline{x_{2}} \lor x_{3}\}$, the set
$\{ 
\overline{x_{1}} \land \overline{x_{2}} \land \overline{x_{3}},
\overline{x_{1}}  \land \overline{x_{2}} \land x_{3},
\overline{x_{1}} \land x_{2} \land \overline{x_{3}},
\overline{x_{1}} \land x_{2} \land x_{3},
x_{1} \land x_{2} \land \overline{x_{3}},
x_{1} \land x_{2} \land \overline{x_{3}}\}$

are the 8-2=6 solutions for $K=\{k_1\}$ set.

\item $k_{2} = {x_{2}} \lor {x_{3}}$   
$\Rightarrow$ $k_{2}^{full}=\{ \overline{x_{1}} \lor {x_{2}} \lor {x_{3}},
x_{1} \lor {x_{2}} \lor {x_{3}}\}$
$\Rightarrow$
set $\{<1,0,0>,<0,0,0>\}$ of unsatisfying variations 
$\Rightarrow$
unsatisfiability pattern $P_{2} = <-,0,0>$  $\Rightarrow$ 2 unsatisfying variations (also 2 clauses in $k_{2}^{full}$).

To calculate the cardinality of the union of two patterns (sets of variations) $P_{1}$ and $P_{2}$, then the intersection of both sets is required: $|P_{1} \cup P_{2}|=|P_{1}|+|P_{2}|-|P_{1} \cap P_{2}|$. 
The intersection of two patterns is obtained by merging them. If there are complementary values for the same position, then the intersection is empty.

$\Rightarrow$ 
$P_{12} = P_{1} \cap P_{2} = intersect(<1,0,->,<-,0,0>)=<1,0,0>$, whose cardinality is $2^{0}=1$. It is easy to check that $<1,0,0>$ is the only element appearing in both sets of unsatisfying variations of $k_{1}$ and $k_{2}$ (and also
that only the clause $\overline{x_{1}} \lor {x_{2}} \lor x_{3}$ appears in both sets $k_{1}^{full}$ and $k_{2}^{full}$)

By the moment the number of unsatisfying variations for $k_{1}$ and $k_{2}$ is $2+2-1=3$, and the number of variations that satisfy both clauses is $2^{3}-3=5$, as can be checked in Table~\ref{truthTableK12}

\begin{table}[h]
	\centering
	\begin{tabular}{|c|c|c||c|c||c|}		
		\hline   $x_{1}$ & $x_{2}$  &   $x_{3}$ & $k_{1}$ & $k_{2}$ & $K=\{k_1,k_2\}$ \\
		\hline  0 & 0  & 0 & 1 & 0 & 0 \\		
		\hline  \cellcolor{blue!25}0 & \cellcolor{blue!25}0  &  \cellcolor{blue!25}1 & 1 & 1&\cellcolor{blue!25}1\\		
		\hline   \cellcolor{blue!25}0 & \cellcolor{blue!25}1  & \cellcolor{blue!25}0 & 1 & 1 & \cellcolor{blue!25}1\\		
		\hline   \cellcolor{blue!25}0 & \cellcolor{blue!25}1  & \cellcolor{blue!25}1 & 1 & 1 & \cellcolor{blue!25}1\\		
		\hline   1 & 0  & 0 & 0 &  0 & 0\\		
		\hline   1 & 0  & 1 & 0 &  1 & 0\\		
		\hline   \cellcolor{blue!25}1 &  \cellcolor{blue!25}1  &  \cellcolor{blue!25}0 & 1 & 1 & \cellcolor{blue!25}1\\		
		\hline   \cellcolor{blue!25}1 & \cellcolor{blue!25}1  & \cellcolor{blue!25}1 & 1 & 1 & \cellcolor{blue!25}1\\	
		\hline	
	\end{tabular} 
	\caption{Truth table for clauses $k_1$ and $k_{2}$}
	\label{truthTableK12}
\end{table} 

\item $k_{3} = \overline{x_{2}} \lor \overline{x_{3}}$ 
$\Rightarrow$ $k_{3}^{full}=\{ \overline{x_{1}} \lor \overline{x_{2}} \lor \overline{x_{3}},
x_{1} \lor \overline{x_{2}} \lor \overline{x_{3}}\}$
$\Rightarrow$  
unsatisfiability pattern $P_{3} = <-,1,1>$  $\Rightarrow$ 2 variations/clauses

$\Rightarrow$ 
$P_{13} = P_{1} \cap P_{3} = intersect(<1,\underline{0},->,<-,\underline{1},1>)=\emptyset$, because the values \underline{0} and \underline{1} appear in the position corresponding to the variable $x_{2}$.

$\Rightarrow$ 
$P_{23} = P_{2} \cap P_{3} = intersect(<-,0,0>,<-,1,1>)=\emptyset$

At this point $P_{1} \cap P_{2} \cap P_{3}$ is also needed because $|P_{1} \cap P_{2} \cap P_{3}|$ has to be added now (by following the formula presented in section \ref{countClauses}). Fortunately $P_{1} \cap P_{2} \cap P_{3}$ can be obtained by using the results of previously calculated intersections. In this case, $P_{123} = P_{1} \cap P_{2} \cap P_{3} = (P_{1} \cap P_{2}) \cap P_{3} = P_{12} \cap P_{3}$

$\Rightarrow$ $P_{123} = P_{12} \cap P_{3} = intersect(<1,0,0>,<-,1,1>)=\emptyset$

Unsatisfying variations now are 3 + 2 - 0 - 0 + 0 = 5 

\item $k_{4} = {x_{1}} \lor \overline{x_{2}} \lor {x_{3}}$ 
$\Rightarrow$ $k_{4}^{full}=\{ {x_{1}} \lor \overline{x_{2}} \lor {x_{3}}\}$
$\Rightarrow$  
pattern $P_{4} = <0,1,0>$  $\Rightarrow$ 1 variation/clause

$\Rightarrow$ 
$P_{14} = P_{1} \cap P_{4} = intersect(<1,0,->,<0,1,0>)=\emptyset$

$\Rightarrow$ 
$P_{24} = P_{2} \cap P_{4} = intersect(<-,0,0>,<0,1,0>)=\emptyset$

$\Rightarrow$ 
$P_{34} = P_{3} \cap P_{4} = intersect(<-,1,1>,<0,1,0>)=\emptyset$

The only other intersection that needs to be calculated is $P_{12} \cap P_{4}$ because all the other ones are empty, so the intersection with $P_{4}$ is also empty.

$P_{124} = P_{12} \cap P_{4} = intersect(<1,0,0>,<0,1,0>)=\emptyset$

Unsatisfying variations now are 5 + 1 - 0 - 0 - 0 + 0 = 6, and as there are no more clauses in $K_1$ then $K_1$ is satisfiable and has 2 solutions. Those solutions are the complementaries of $(K_1^{full})^C$ 
$=\{ 
x_{1} \lor {x_{2}} \lor \overline{x_{3}},
\overline{x_{1}} \lor \overline{x_{2}} \lor x_{3}\}$.
Although this algorithm does not provide any of those solutions, the algorithm presented in section~\ref{solAlg} will be able to do it. 
\end{itemize}


Let us continue with the rest of clauses of the unsatisfiable set
$K_2$

\begin{itemize}
\item $k_{5} = {x_{1}} \lor {x_{2}}$
$\Rightarrow$ $k_{5}^{full}=\{ {x_{1}} \lor {x_{2}} \lor \overline{x_{3}},
{x_{1}} \lor {x_{2}} \lor {x_{3}}\}$
$\Rightarrow$  
pattern $P_{5} = <0,0,->$  $\Rightarrow$ 2 variations/clauses

$P_{15} = P_{1} \cap P_{5} = intersect(<1,0,->,<0,0,->)=\emptyset$

$\Rightarrow$ 
$P_{25} = P_{2} \cap P_{5} = intersect(<-,0,0>,<0,0,->)=<0,0,0>$, 1 variation to substract

$\Rightarrow$ 
$P_{35} = P_{3} \cap P_{5} = intersect(<-,1,1>,<0,0,->)=\emptyset$

$\Rightarrow$ 
$P_{45} = P_{4} \cap P_{5} = intersect(<0,1,0>,<0,0,->)=\emptyset$

$\Rightarrow$ 
$P_{125} = P_{12} \cap P_{5} = intersect(<1,0,0>,<0,0,->)=\emptyset$

Unsatisfying variations now are 6 + 2 - 0 - 1 - 0 - 0 + 0 = 7

\item $k_{6} = \overline{x_{1}} \lor \overline{x_{2}}$
$\Rightarrow$ $k_{6}^{full}=\{ \overline{x_{1}} \lor \overline{x_{2}}  \lor \overline{x_{3}},
\overline{x_{1}} \lor \overline{x_{2}}  \lor {x_{3}}\}$
$\Rightarrow$  
pattern $P_{6} = <1,1,->$  $\Rightarrow$ 2 variations/clauses

$P_{16} = P_{1} \cap P_{6} = intersect(<1,0,->,<1,1,->)=\emptyset$

$\Rightarrow$ 
$P_{26} = P_{2} \cap P_{6} = intersect(<-,0,0>,<1,1,->)=\emptyset$

$\Rightarrow$ 
$P_{36} = P_{3} \cap P_{6} = intersect(<-,1,1>,<1,1,->)=<1,1,1>$, 1 variation to substract

$\Rightarrow$ 
$P_{46} = P_{4} \cap P_{6} = intersect(<0,1,0>,<1,1,->)=\emptyset$

$\Rightarrow$ 
$P_{56} = P_{5} \cap P_{6} = intersect(<0,1,0>,<1,1,->)=\emptyset$

$\Rightarrow$ 
$P_{126} =P_{12} \cap P_{6} = intersect(<1,0,0>,<1,1,->)=\emptyset$

$\Rightarrow$ 
$P_{256} = P_{25} \cap P_{6} = intersect(<0,0,0>,<1,1,->)=\emptyset$

Unsatisfying variations now are 7 + 2 - 0 - 0 - 1 - 0 - 0 + 0 + 0 = 8. Therefore, with this clause the set of clauses is unsatisfiable
\end{itemize}

Although there are potentially many variations of patterns that require to be calculated, just when an intersection among some variations is empty,
then adding more patterns to intersect with that is also empty, and they do not need to be calculated. In this case, from all the possible intersections among the patterns $P_{1}, P_{2}, P_{3}, P_{4}, P_{5}$ and $P_{6}$, only 3 were not empty: $P_{12}$, $P_{25}$ and $P_{36}$, that is 3 of 57 (${6 \choose 2} + {6 \choose 3} + {6 \choose 4} + {6 \choose 5} + {6 \choose 6}=15+20+15+6+1=57$). The reason for this is that the clauses are relatively dense: the proportion among the number of literals and the number of variables is relatively high. In this case that proportion is 1 for clause $k_{4}$ (3 literals, 3 variables) and 0.66 for the rest of clauses (2 literals, 3 variables). Moreover, when the intersection among 
patterns is not empty, the result pattern is even more dense: $P_{12}, P_{25}$ and $P_{36}$ have a density of 1.

This is exactly the advantage mentioned in section~\ref{Intro}: having many literals in the clauses (big values of $k$) increments the possibilities of finding complementary literals among the clauses what results in empty intersections. In this paper, we will see that this algorithm can be executed in polynomial time for all instances of $k$-SAT with relatively {\it dense} clauses; clauses where $k$ is relatively close to $n$ ($k\geq7\sqrt{n}$ or even $k\geq5\sqrt{n}$).

\section{Previous concepts employed by the algorithm}

Once the algorithm has been explained with a motivating example, the definitions and proofs of some concepts are going to be presented in this section.

\subsection{The sets $K^{*}$, $K^{full}$, $\overline{K}^{full}$ and $\overline{K}$}
\label{definitions} 

Let $K$= \{$k_{1}\cdots k_{m}$\} be a CNF formula of $m$ disjunctive clauses with literals of $n$ variables $x_{1} \cdots x_{n}$. The next sets can be defined:

\begin{itemize}
	\item $K^{*}$ is the set of all posible disjunctive clauses that can be built with the {\it different} variations of literals of the $n$ variables.
	
	$K^{*} =\{(x_{1},\cdots,x_{n}),(x_{1},\cdots,\overline{x_{n}}),\cdots, (\overline{x_{1}},\cdots,\overline{x_{n}})\}$ 
	
	\begin{itemize}
	
	 \item Each clause in $K^{*}$ is a {\it full} clause that contains exactly one literal for each variable.
	  
	 \item $K^{*}$ contains $2^{n}$ different full clauses: $|K^{*}|=2^n$
	
	\end{itemize}

\item $K^{full}$ contains all the full disjunctive clauses corresponding to the clauses $k_{i}$ of $K$:

$K^{full} = \bigcup_{i=1}^{m}(k_{i}^{full})$

where $k_{i}^{full}$ is the set of all {\it full} clauses obtained by adding to the clause $k_{i}$ with $j$ literals every variation of the $x^i_{j+1},\cdots,x^i_{n}$ variables without literal in $k_{i}$

$k_i^{full}$ = \{$k_i \lor x^i_{j+1} \lor \cdots \lor x^i_{n}$,
$k_i \lor x^i_{j+1} \lor \cdots \lor \overline{x^i_{n}}$,$\cdots$,
$k_i \lor \overline{x^i_{j+1}} \lor \cdots \lor \overline{x^i_{n}}$\}
	
	\item $\overline{K}^{full}$ contains all the full conjunctive clauses that are the complementaries of the clauses in $K^{full}$:
	
	$\overline{K}^{full} = \{c |c=\overline{d} \land d\in K^{full} \}$  
	
	For each clause in $\overline{K}^{full}$ there is a clause in $K^{full}$, then $|\overline{K}^{full}|=|K^{full}|$ 
	
	\item $\overline{K}$ contains all the conjunctive clauses that are the complementaries of the clauses $k_{i}$ of $K$:

$\overline{K} = \{c |c=\overline{d} \land d\in K \}$  

\end{itemize}

\subsection{$K^{*}$ is unsatisfiable}
\label{allClausesAreUnsatisfiable}

Let $x_{1} \cdots x_{n}$ be a set of variables, then the conjuction of all the possible (disjunctive) clauses that can be formed with their literals is False, and therefore, unsatisfiable:  $(x_{1} \lor \cdots \lor x_{n}) \land (x_{1} \lor \cdots \lor \overline{x_{n}}) \land \cdots \land (\overline{x_{1}} \lor \cdots \lor \overline{x_{n}}) = False$

\vspace{0.3cm}
{\bf Proof by induction}

\begin{itemize}
\item It is true for $n=1$ because $(x_{1} \land\overline{x_{1}})=False$

\item We suppose it is true for $n=k$

$(x_{1} \lor \cdots \lor x_{k}) \land (x_{1} \lor \cdots \lor \overline{x_{k}}) \land \cdots \land (\overline{x_{1}} \lor \cdots \lor \overline{x_{k}}) = False$

\item Is it also true for  $n=k+1$?

$(x_{1} \lor \cdots \lor x_{k} \lor x_{k+1}) \land (x_{1} \lor \cdots \lor x_{k} \lor \overline{x_{k+1}}) \land \cdots \land (\overline{x_{1}} \lor \cdots \lor \overline{x_{k}} \lor x_{k+1}) \land (\overline{x_{1}} \lor \cdots \lor \overline{x_{k}} \lor \overline{x_{k+1}})=$

$((x_{1} \lor \cdots \lor x_{k}) \land (x_{1} \lor \cdots \lor \overline{x_{k}}) \land \cdots \land (\overline{x_{1}} \lor \cdots \lor \overline{x_{k}})) \lor x_{k+1}) \land$

$((x_{1} \lor \cdots \lor x_{k}) \land (x_{1} \lor \cdots \lor \overline{x_{k}}) \land \cdots \land (\overline{x_{1}} \lor \cdots \lor \overline{x_{k}})) \lor \overline{x_{k+1}})=$

$(False \lor x_{k+1}) \land (False \lor \overline{x_{k+1}}) = 
(x_{k+1}) \land (\overline{x_{k+1}}) = False$

\item Yes, it is true for $n=k+1$, and therefore the proposition is proven.

\end{itemize}

\subsection{$K$ is equivalent to its expanded set of full clauses $K^{full}$}
\label{KKfullequiv}

Now the following two properties can be proven:

\begin{itemize}
	
\item

The set $\{k\}$ with a disjunctive clause $k$ is equivalent to $k^{full}$

As any clause $k$ is equivalent to $(k \lor False)$, and the conjunction of all possible variation of clauses formed with a set of variables is $False$ (see section~\ref{allClausesAreUnsatisfiable})

$k = (k \lor False) =$ 

$=(k \lor ((x_{i+1} \lor \cdots \lor x_{n}) \land (x_{i+1} \lor \cdots \lor \overline{x_{n}}) \land \cdots \land (\overline{x_{i+1}} \lor \cdots \lor \overline{x_{n}}))=$

$=((k \lor x_{i+1} \lor \cdots \lor x_{n}) \land (k \lor x_{i+1} \lor \cdots \lor \overline{x_{n}}) \land \cdots \land (k \lor \overline{x_{i+1}} \lor \cdots \lor \overline{x_{n}}))$ 

$\Rightarrow \{k\} \equiv  k^{full}$

\item 
 $K$ is equivalent to $K^{full}$

$K = \{k_{1}, \cdots k_{m}\}= \bigcup_{i=1}^{m}(\{k_{i}\}) =  \bigcup_{i=1}^{m}(k_{i}^{full}) = K^{full}$

\end{itemize}

\subsection{$K$ is unsatisfiable when its $K^{full}$ contains $2^n$ clauses }
\label{fullContainsAllClauses}

 $K^{*}=\{(x_{1},\cdots,x_{n}),(x_{1},\cdots,\overline{x_{n}}),\cdots, (\overline{x_{1}},\cdots,\overline{x_{n}})\}$ is the set of all posible and {\it different} variations of clauses that can be formed with $n$ variables where $|K^{*}|=2^n$ (according to ~\ref{definitions}).
If the $K^{full}$ set derived from a set $K$ also contains $2^n$ different clauses, 
$|K^{full}|=2^n$, then they must be all the clauses in $K^{*}$:

$\Rightarrow$ 
$K^{full} =K^{*}$ 

$\Rightarrow$ 
$K^{full}$ is unsatisfiable (according to ~\ref{allClausesAreUnsatisfiable})

$\Rightarrow$ 
$K$ is unsatisfiable (according to ~\ref{KKfullequiv})

\subsection{Complementaries of clauses in $(K^{full})^{C}$ satisfy $K$}
\label{solutionsForK}

Once known that $K$ is unsatisfiable when $K^{full} =K^{*}$, we analyze now what happens when 
$K^{full} \subset K^{*}$.  Let $(K^{full})^{C}$ be the set of 
disjunctive full clauses that are in $K^{*}$ but not in 
$K^{full}$, that is $(K^{full})^{C}=K^{*}-K^{full}$. Notice that $K^{full} \cup (K^{full})^{C}=K^{*}$, what means that $K^{full}$ and $(K^{full})^{C}$ are complementary sets and $c \in (K^{full})^{C}$ 
$\leftrightarrow$ $c \notin K^{full}$.
In this case, the complementaries of 
clauses of $(K^{full})^{C}$ are solutions that satisfy $K$:

\begin{center}
$c= (c_{1},\cdots,c_{n}) = (c_{1}\lor\cdots \lor c_{n}) \in (K^{full})^{C}$ 
$\rightarrow$ 
 
$\rightarrow$ $\overline{c}= (\overline{c_{1}},\cdots,\overline{c_{n}}) 
= (\overline{c_{1}}\land \cdots \land \overline{c_{n}})$ is a solution of $K$
\end{center}

\vspace{0.3cm}
{\bf Proof by induction}

\begin{itemize}
	\item For $n=1$, these are all the possibilites with elements of  $K^{*}_{1}=\{x_{1},\overline{x_{1}}\}$:
	
	\begin{enumerate}
		\item $K$=\{$x_{1}$\} $\Rightarrow$ $K^{full}$=\{$x_{1}$\}	
		$\Rightarrow$ $\overline{x_{1}} \in (K^{full})^{C}$
		$\Rightarrow$ $\overline{\overline{x_{1}}}=x_{1}$ is a solution for $K$=\{$x_{1}$\}
		
		\item $K$=\{$\overline{x_{1}}$\} $\Rightarrow$ $K^{full}$=\{$\overline{x_{1}}$\}	
		$\Rightarrow$ $x_{1} \in (K^{full})^{C}$
		$\Rightarrow$ $\overline{x_{1}}$ is a solution for $K$=\{$\overline{x_{1}}$\}	
		
		\item $K$=\{$x_{1}$,$\overline{x_{1}}$\} $\Rightarrow$ $K^{full}$=\{$x_{1}$,$\overline{x_{1}}$\}	
		$\Rightarrow$ 
		$(K^{full})^{C}=\emptyset$ 
		$\Rightarrow$
		$\nexists c \in (K^{full})^{C}$ 
	\end{enumerate}
	Therefore, in all cases for $n=1$ \{$c \in (K^{full})^{C}$ 
	$\rightarrow$ $\overline{c}$ is a solution of $K$\} is true 
	
	\item We suppose that it is true for $n=q$
	
	$c= (c_{1},\cdots,c_{q}) \in (K^{full})^{C}$ 
	$\rightarrow$ $\overline{c}= (\overline{c_{1}},\cdots,\overline{c_{q}})$ is a solution of $K$ (Result 1)
	
	\item Is it also true for  $n=q+1$?
	
	{\bf Proof by reductio ad absurdum}
	
	Let us suppose that it is false:
	
	$((c_{1},\cdots,c_{q+1}) \in (K^{full})^{C}$ 
	$\rightarrow$ $(\overline{c_{1}},\cdots,\overline{c_{q+1}})$ is a solution of $K$) = $False$
	
	$\Rightarrow$
	$(c_{1},\cdots,c_{q+1}) \in (K^{full})^{C}$ 
	$\land$ $(\overline{c_{1}},\cdots,\overline{c_{q+1}})$ is {\bf not} a solution of $K$
	
	$\Rightarrow$
	$(c_{1},\cdots,c_{q},c_{q+1}) \in (K^{full})^{C}$ 
	$\land$ $(\overline{c_{1}},\cdots,\overline{c_{q}},\overline{c_{q+1}})$ is {\bf not} a solution of $K$
	
	$\Rightarrow$
	$(c_{1}\lor\cdots\lor c_{q} \lor c_{q+1}) \in (K^{full})^{C}$
	(Result 2)

	$\land$ $(\overline{c_{1}},\cdots,\overline{c_{q}},\overline{c_{q+1}})$ is {\bf not} a solution of $K$
	(Result 3)
	
	Let us evaluate the clauses of $K^{full}$ with $(\overline{c_{1}},\cdots,\overline{c_{q}},\overline{c_{q+1}})$ then:
	
	\begin{itemize}
		\item 
		The literal $\overline{c_{q+1}}$ will satisfy all the clauses in  $K^{full}$ containing literal $\overline{c_{q+1}}$, because they are conjunctive clauses ($k' \lor \overline{c_{q+1}} = k' \lor True=True$). Let $K_{rest}^{full}$ be
		the rest of clauses to be satisfied. 
		
		Notice that if the clause $(c_{1}\lor\cdots\lor c_{q} \lor \overline{c_{q+1}})$ were in $K^{full}$, then  
		$(c_{1}\lor\cdots\lor c_{q})$ would not be in $K_{rest}^{full}$.
		
		\item
		The literal $\overline{c_{q+1}}$ will eliminate the literal $c_{q+1}$ in the clauses of $K_{rest}^{full}$ that contain it
		($k' \lor c_{q+1} = k' \lor False = k'$).
		As the previous result 2 says that $(c_{1}\lor\cdots\lor c_{q} \lor c_{q+1}) \in (K^{full})^{C}$ then $(c_{1}\lor\cdots\lor c_{q} \lor c_{q+1}) \notin K^{full}$, and therefore 
		$(c_{1}\lor\cdots\lor c_{q})$ will not be either in $K_{rest}^{full}$.
		
		\item
		At this point it is known that clause 
		$(c_{1}\lor\cdots\lor c_{q})$ $\notin$ $K_{rest}^{full}$
		and that $K_{rest}^{full}$ is composed of $q$ variables and not $q+1$ variables (because literals $c_{q+1}$ and $\overline{c_{q+1}}$ do not appear in clauses of $K_{rest}^{full}$). As the proposition for $n=q$ is true (Result 1), then 
		$(\overline{c_{1}},\cdots,\overline{c_{q}})$ is a solution for $K_{rest}^{full}$
		
		\item
		It the literal $\overline{c_{q+1}}$ has satisfied part of $K^{full}$ reducing it to $K_{rest}^{full}$, and the literals $(\overline{c_{1}},\cdots,\overline{c_{q}})$ have satisfied $K_{rest}^{full}$, then $(\overline{c_{1}},\cdots,\overline{c_{q}},\overline{c_{q+1}})$ is a solution for K, what it is a contradiction that demonstrates that the proposition for $n=q+1$ is also true.

	\end{itemize}

\end{itemize}

Therefore the solutions that satisfy $K$ are the complementaries of clauses in $(K^{full})^{C}$, or what it is the same, the conjunctive clauses that are in
$(\overline{K}^{full})^{C}$.

\subsection{Counting clauses in $K^{full}$/$\overline{K}^{full}$ directly from $K$/$\overline{K}$}
\label{countClauses}

To build the $K^{full}$ set derived from $K$ and find if there are solutions (complementaries of full clauses that are in $K^{*}$ but not in $K^{full}$) that make $K$ satisfiable leads to an $O(2^n)$ algorithm. It is obvious that if $K$ is unsatisfiable then $K^{full}$ would be $K^{*}$, that contains $2^n$ full clauses.
It is much more efficient to count the number of clauses that would be in $K^{full}$ directly from $K$, which have been shown to be equivalent to $K^{full}$ in section~\ref{KKfullequiv}. In that case, the number of solutions would be ($2^n -|K^{full}|$)  

In order to calculate the size of those sets of clauses, then common formula of combinatorics such as calculating the number of variations and the inclusion-exclusion principle \cite{Sane2013} can be used.	

\begin{itemize}
	\item The number of full clauses corresponding to a clause $k$ with $j$ literals corresponding to $n$ different variables is $2^{n-j}$, that is the number of variations that can be obtained with two literals $x_{p}$ or $\overline{x_{p}}$ for the ($n-j$) remaining variables, where $p\in\{j+1,n\}$. Therefore, $|\{k\}|=2^{n-j}$
	
	\item
	
	The number of full clauses in $K=K^{full}=\bigcup_{i=1}^{m}(k_{i}^{full})=\bigcup_{i=1}^{m}(\{k_{i}\})$ 
	is determined by the principle of inclusion-exclusion:

	$|K| = |\bigcup_{i=1}^{m}(\{k_{i}\})| =$ 	
	$\sum_{i=1}^{m} |\{k_{i}\}|
	-\sum_{j,k: 1\leq j<k\leq m} |\{k_{j}\}\cap \{k_{k}\}|+$

	$+\sum_{j,k,l: 1\leq j<k<l\leq m} |\{k_{j}\}\cap \{k_{k}\} \cap \{k_{l}\}|$	
	$+\cdots+ (-1)^{m+1}|\{k_{1}\}\cap \cdots \cap \{k_{m}\}|$
	
\end{itemize}

Due to the way in which $\overline{K}^{full}$ set has been built (see section \ref{definitions}), then $|K^{full}|=|\overline{K}^{full}|$ and, with a similar reasoning, the following can be concluded:

\begin{itemize}
\item $|\{\overline{k}\}|=2^{n-j}$ 

\item $|\overline{K}| = |\bigcup_{i=1}^{m}(\{\overline{k_{i}}\})| =$ 	
$\sum_{i=1}^{m} |\{\overline{k_{i}}\}|
-\sum_{j,k: 1\leq j<k\leq m} |\{\overline{k_{j}}\}\cap \{\overline{k_{k}}\}|+$

$+\sum_{j,k,l: 1\leq j<k<l\leq m} |\{\overline{k_{j}}\}\cap \{\overline{k_{k}}\} \cap \{\overline{k_{l}}\}|$	
$+\cdots+ (-1)^{m+1}|\{\overline{k_{1}}\}\cap \cdots \cap \{\overline{k_{m}}\}|$
\end{itemize}
\subsection{Intersections of conjunctive clauses are efficient}
The intersection of clauses $c$ and $d$ is the set of clauses built with the variations of literals of variables that satisfy  $c$ and that also satisfy $d$: $\{c\}\cap \{d\} = \{c \land d\}$. When the clauses are conjunctive ($\overline{k_{i}}$ and $\overline{k_{j}}$ are conjunctive) then the intersection is also a conjunctive clause that is much more efficient to calculate\footnote{If the clauses are
	disjunctive, the intersection clause is not disjunctive:
	$k_{i}\land k_{j} =
	(l_{i1}\land l_{j1}) \lor (l_{i1}\land l_{j2}) \lor \cdots (l_{ip}\land l_{jq})$}

\begin{center}
	$\overline{k_{i}}\land \overline{k_{j}} = $
	
	$= (\overline{l_{i1}} \land \overline{l_{i2}}\cdots\land \overline{l_{ip}}) \land (\overline{l_{j1}} \land \overline{l_{j2}}\cdots\land \overline{l_{jq}})= (\overline{l_{i1}} \land \overline{l_{i2}}\cdots\land \overline{l_{ip}} \land \overline{l_{j1}} \land \overline{l_{j2}}\cdots\land \overline{l_{jq}})$
\end{center}

The good news here is that if there are complementary literals ($\exists r,s: l_{jr}=\overline{l_{is}}$ ), then $\overline{k_{i}}\land \overline{k_{j}} = \emptyset$. Moreover, any other intersection with other clauses including $\overline{k_{i}}$ and $\overline{k_{j}}$ is also empty because $k \land \emptyset = \emptyset$ for any $k$.
As it is much more efficient to calculate intersections of conjunctive clauses, then it will be much better to calculate $|\overline{K}|$ instead of $|K|$

\section{SARRIGUREN: a complete algorithm for SAT}

The algorithm returns $(2^n-|\overline{K}|)$, the number  of solutions or satisfying assignments
of a set $K$ of $m$ disjunctive clauses $\{k_{1}, \cdots k_{m}\}$ with $n$ variables. 
$K$ is unsatisfiable if there are zero solutions, and satisfiable if there at least one.
The value $|\overline{K}|$ is calculated 
by applying the formula
$\sum_{i=1}^{m} |\{\overline{k_{i}}\}|$ 
$-\sum_{j,k} |\{\overline{k_{j}}\}\cap \{\overline{k_{k}}\}|$
$+\cdots (-1)^{m+1}|\{\overline{k_{1}}\}\cap \cdots \cap \{\overline{k_{m}}\}|$.
The cardinality of all the conjunctive clauses $|\{\overline{k_{i}}\}|$ and intersecting clauses $|\{\overline{k_{1}}\}\cap \cdots \cap \{\overline{k_{m}}\}|$
is calculated as $2^{n-var\#}$, and the intersection of conjunctive clauses is the result of merging the literals of the clauses (the intersection is $\emptyset$ if clauses contain complementary literals).
The cardinalities are calculated 
in this order:
$+|\{\overline{k_{1}}\}|$, $+|\{\overline{k_{2}}\}|$, 
$-|\{\overline{k_{1}}\}\cap \{\overline{k_{2}}\}|$, 
$+|\{\overline{k_{3}}\}|$, 
$-|\{\overline{k_{1}}\}\cap \{\overline{k_{3}}\}|$,
$-|\{\overline{k_{2}}\}\cap \{\overline{k_{3}}\}|$,
$+|\{\overline{k_{1}}\}\cap \overline{k_{2}}\}\cap \{\overline{k_{3}}\}|$,
$\cdots$
$+/-|\{\overline{k_{1}}\}\cap \cdots \cap \{\overline{k_{m}}\}|$. Whenever an intersection of clauses is empty, then it is not further processed with other clauses. And, after the processing of a clause $k_j$ (by calculating cardinalities with previous clauses  $k_{1} \cdots k_{j-1}$, if $|\overline{K}|=2^n$, then the algorithm ends by returning zero. Finally, $k_i$ can be used instead of $\overline{k_{i}}$, because the counting of clauses and calculation of intersections gets the same results with both of them.

\begin{algorithm}[H]
	\caption{SARRIGUREN algorithm}
\begin{algorithmic}
	\State {\bf Input:} $K = k_{1} \cdots k_{m}$, a set of disjuntive clauses sorted by variable number
	\State 	\hspace{1.3cm}$n$: number of variables 
	
	\State {\bf Output:} 0 if $K$ is {\tt UNSAT} or the number of solutions if it is {\tt SAT}
	 
	\State  {\bf Precondition:}  there are no complementary literals in any clause 
	\State $P \gets$ [ ]  
	\hspace{1cm} \Comment{$P$ contains the signed patterns processed until now}
	\State $u \gets 0$ 
	\hspace{1cm} \Comment{$u$ number of unsatisfiable variations at the moment}
	\State \For{$i$ in 1..$n$}
	\State $ap \gets$ {\bf pattern}($k_{i}$) 
	\hspace{1.5cm} \Comment{$ap$ is the pattern of the actual clause $k_{i}$}
	\State $N \gets$ [ ] 
	\hspace{1cm} \Comment{$N$: signed patterns to process with clauses $k_{i+1}\cdots k_{n}$}
	\State $u \gets u +$ {\bf cardinality}($ap$,$n$) 
	\hspace{0.5cm} \Comment{add to $u$ the \#{\tt unsat} variations of $k_{i}$}
	\State {\bf add} $<$'-',$ap$)$>$ {\bf to} $N$ 
	\hspace{0.25cm} \Comment{adds the negative pattern of $k_{i}$ in order to}
	\State \hspace{3.5cm} \Comment{substract repeated variations in $k_{i+1}\cdots k_{n}$}
	\State \For{$<sign,p>$ in $P$}
	\hspace{0.2cm} \Comment{signed patterns processed in $k_{1} \cdots k_{i-1}$}
	\State \hspace{0.1cm} \Comment{$p$ is the pattern to process with $ap$, the $sign$ is '+' or '-'}
	\State $ip$ $\gets$ {\bf intersect}($ap$,$p$) \Comment{$ip$ is the intersection pattern}
	\If{$ip$ is not empty} 
	\State $u \gets u$ $sign$ {\bf cardinality}($ip$,$n$) 
	\hspace{0.2cm} \Comment{\#{\tt unsat} variations of the} 
	\State \hspace{3cm}\Comment{intersection are added or substracted}
	\State {\bf add} $<${\bf contrary}($sign$),$ip$$>$ {\bf to} $N$
	\Comment{adds intersection pattern}
	\State \Comment{with the contrary sign to add/substract repeated in $k_{i+1}\cdots k_{n}$}
		
	\EndIf
	\EndFor	
	\State {\bf append} $N$ {\bf to} $P$
	\hspace{0.25cm} \Comment{the new signed patterns calculated by processing $k_{i}$} 
	\State \hspace{2.9cm} \Comment{are added in order to be processed with $k_{i+1}\cdots k_{n}$}
	\If{$u$ = $2^{n}$} 
		\State {\bf return} 0 \hspace{0.3cm} \Comment{{\tt UNSAT:} {\it all the variations are unsatisfiable}}
		\EndIf
	\EndFor		 
	
	\State {\bf return} ($(2^{n} - u)$) \hspace{1.7cm} \Comment{{\tt SAT:} {\it there are $(2^{n} - u)$ solutions}} 
	\State \hrulefill
	\vspace{0.1cm}
	\State {\bf pattern} ($c$) returns $c$ {\it\{And not $\overline{c}$ because {\bf \footnotesize cardinality} and {\bf \footnotesize intersect} work equal\}}
	\State {\bf cardinality} ($c$,$n$) returns $2^{n-nv}$, where $nv$ is the number of literals in $c$
	\State {\bf contrary} ($s$) returns + if $s$ is -, and - if $s$ is +
	
	\State {\bf intersect} ($c$,$d$) returns sorted merge of $c$ and $d$ or $\emptyset$ if complementary literals

\end{algorithmic}		
\end{algorithm}

Notice that, on the one hand, this algorithm is designed to know if a set is satisfiable, but
not to know which are the solutions that satisfy that set. 
The solutions are the variations that are in $(\overline{K}^{full})^C$. However $\overline{K}^{full}$ is not being built, only counted. Section~\ref{solAlg} presents an algorithm that obtains solutions by using SARRIGUREN. On the other hand, SARRIGUREN is an algorithm for \#SAT and for Unique-SAT because it counts the total number of satisfying assignmens, and because it can be  easily checked if there is only one or not.

\section{Analysis of complexity of SARRIGUREN}

In the algorithm there is a loop ({\bf for} {\it i} in 1..$n$ {\bf do})  that processes each clause $k_{i}$ in this way: it adds $|k_{i}|$ to a counter, it calculates the intersections between $k_{i}$ and all the clauses $p$ stored in $P$ ({\bf intersect}($ap$,$p$)), where $P$ contains all the previous clauses $k_{1} \cdots k_{i-1}$, and all the non-empty or overlapping intersections among clauses $k_{1} \cdots k_{i-1}$. While those intersections are calculated then their sizes $|k_{i} \cap k_{x} \cap \cdots \cap k_{y}|$ are added or deleted from the counter, according to the formula of section \ref{countClauses}. The new overlapping intersections found in each iteration are added to $N$ ({\bf add} $<${\bf contrary}($sign$),$ip$$>$ {\bf to} $N$), and after the processing of clause $k_{i}$, added to $P$ ({\bf append} $N$ {\bf to} $P$) to be processed with clause $k_{i+1}$ in the next iteration of the loop.

The key point in order to analyze the complexity of that algorithm is the number of overlapping intersections of clauses  
$\overline{k_{i}}$ and $\overline{k_{j}}$. If that number of overlapping intersections is close to zero, then the complexity of the algorithm decreases substantially because no new intersection clauses are added to $P$.

In the following the probability of overlapping intersections clauses is first analyzed, then the number of new clauses that need to be processed. After that, the analysis of the complexity of SARRIGUREN algorithm (applied to $k$-SAT with dense clauses) for the average case will be discussed, and also some remarks for the best and worst cases.

\subsection{Probability of overlapping among dense clauses}

Let us find the probability of the intersection of clauses $c$ and $d$ to overlap (or not to be disjoint). That happens when there are no complementary literals in clauses $c$ and $d$. If a clause $c$ has $k_{c}$ literals corresponding to $n$ possible variables, the possibilities of clause $d$ with $k_{d}$ literals and also of $n$ variables not to have a complementary literal with $c$ are these ones:

\begin{itemize}

\item All the $k_{d}$ literals of $d$ are chosen from the $(n-k_{c})$ variables without literal of $c$: $n-k_{c} \choose k_{d}$
combinations where each combination of $k_{d}$ literals can variate with the 2 possibilities for each literal: $2^{k_{d}}$. There are in total ${n-k_{c} \choose k_{d}} \times {2^{k_{d}}}$ possibilities
(or what it is the same: ${n-k_{c} \choose k_{d}-0} \times (2^{k_{d}-0}) \times {k_{c} \choose 0}$)

\item $(k_{d}-1)$ literals of $d$ are chosen from the $(n-k_{c})$ variables without literal of $c$: $n-k_{c} \choose k_{d} -1$
combinations where each combination of $(k_{d}-1)$ literals can variate with the 2 possibilities for each literal: $2^{k_{d}-1}$. The last literal of $d$ is chosen from the $k_{c}$ variables with literal of $c$, but is the same literal (and does not overlap). There are in total ${n-k_{c} \choose k_{d} -1} \times 2^{k_{d}-1} \times k_{c}$ possibilities (or what it is the same: ${n-k_{c} \choose k_{d}-1} \times (2^{k_{d}-1}) \times {k_{c} \choose 1}$)

\item $(k_{d}-2)$ literals of $d$ are chosen from the $(n-k_{c})$ variables without literal of $c$: $n-k_{c} \choose k -2$ combinations where each combination of $(k_{d}-2)$ literals can variate with the 2 possibilities for each literal: $2^{k_{d}-2}$. The other 2 literals of $d$ are chosen from the $k_{c}$ variables with literal of $c$, but they are the same literal (and do not overlap). There are in total ${n-k_{c} \choose k_{d}-2} \times (2^{k_{d}-2}) \times {k_{c} \choose 2}$ possibilities (or what it is the same: ${n-k_{c} \choose k_{d}-2} \times (2^{k_{d}-2}) \times {k_{c} \choose 2}$)

\item The same can be done by chosing $(k_{d}-i)$ literals  of $d$ until $i=(k_{d}-1)$. 
 There are in total ${n-k_{c} \choose k_{d}-i} \times (2^{k_{d}-i}) \times {k_{c} \choose i}$ possibilities for each case.

\item No literal of $d$ is chosen from the $(n-k_{c})$ variables without literal of $c$. All the $k_{d}$ literals of $d$ are chosen from the $k_{c}$ variables with literal of $c$, but they are the same literal (and do not overlap). There are ${k_{c} \choose k_{d}}$ possibilities (or what it is the same: ${n-k_{c} \choose k_{d}-k_{d}} \times (2^{k_{d}-k_{d}}) \times {k_{c} \choose k_{d}}$)

\end{itemize}

In summary there are $\sum_{i=0}^{k_{d}} ({n-k_{c} \choose k_{d}-i} \times (2^{k_{d}-i}) \times {k_{c} \choose i})$ possible $d$ clauses that overlap with a given $c$ clause, from a total of ${n \choose k_{d}} \times 2^{k_{d}}$ possible $d$ clauses. Therefore, the probability of $c$ and $d$ clauses to overlap is: 
\begin{center}

$P_{overlap} = \frac{\sum_{i=0}^{k_{d}} ({n-k_{c} \choose k_{d}-i} \times (2^{k_{d}-i}) \times {k_{c} \choose i})}{{n \choose k_{d}} \times 2^{k_{d}}}$

\end{center}

Let us see what happens in the $k$-SAT case with very dense clauses, that is, when $k$ is very close to $n$, and $k=k_{c}=k_{d}$:

\[ \lim_{k\to n}P_{overlap}=
\lim_{k\to n} \frac{\sum_{i=0}^{k} ({n-k \choose k-i} \times (2^{k-i}) \times {k \choose i})}{{n \choose k} \times 2^{k}}=
\]

\[
=
\frac{\sum_{i=0}^{n} ({n-n \choose n-i} \times (2^{n-i}) \times {n \choose i})}{{n \choose n} \times 2^{n}}
= \frac{({n-n \choose n-n} \times (2^{n-n}) \times {n \choose n})}{{n \choose n} \times 2^{n}}
=\frac{1}{2^{n}}\]

The only feasible term in the summatory $\sum_{i=0}^{n}$ is the corresponding to $i=n$ because the others contain bad combinatorial numbers: ${0 \choose n}\cdots {0 \choose n-1}$.

Therefore, when the $k$-SAT instance to solve has very {\it dense} clauses, the probability of overlapping intersections $P_{overlap}$ is $\frac{1}{2^{n}}$, that is zero for big values of $n$ ($\lim_{n\to \infty} \frac{1}{2^{n}} = 0$), but 
also very close to zero for small values of $n$ ($n >10$, for example).

\subsection{Number of new overlapping intersections to process}

However, although the probability of overlapping clauses $P_{overlap}$ is very close to zero for $k$-SAT instances with very {\it dense} clauses ($k\to n$), the number of possible intersections among $m$ clauses may not be zero:
${m \choose 2}$ intersections of 2 clauses $\{\overline{k_{j}}\}\cap \{\overline{k_{k}}\}$, ${m \choose 3}$ intersections of 3 clauses 
$\{\overline{k_{j}}\}\cap \{\overline{k_{k}}\} \cap \{\overline{k_{l}}\}$, and so on.
	
Therefore, the expected number of new clauses added to P in the algorithm that need to be processed is ${m \choose 2} \times P_{overlap}$ intersections of 2 clauses 
$\{\overline{k_{j}}\}\cap \{\overline{k_{k}}\}$, that may be significative if $P_{overlap}$ is not small enough.  
Those new non-overlapping intersection clauses will be intersected again with all $\{\overline{k_{l}}\}$ clauses. Fortunately, the probability of overlapping for these intersections of 3 clauses $\{\overline{k_{j}}\}\cap \{\overline{k_{k}}\} \cap \{\overline{k_{l}}\}$ will be smaller because the non-overlapping intersections of 2 clauses will be more dense because, in average, the intersections of 2 clauses will have a 50\% more literals than the intersected clauses.

\subsection{Complexity for the average case of $k$-SAT dense}

The complexity of SARRIGUREN algorithm for SAT, where there are clauses of any size is obviously exponential because there are $\sum_{j=1}^{m} {m\choose j} = (2^m -1)$ possible intersections among $m$ clauses. In this section, the complexity of the algorithm is going to be analyzed but, only for the average case of random $k$-SAT with dense clauses. First, it will be 
discussed how dense have to be the clauses so that their intersections do not overlap, and then, the complexity of the algorithm for such dense clauses will be analyzed.

\subsubsection{Dense clauses do not overlap for the average case}
\label{clausesDoNotOverlap}

The formula $P_{overlap}$ used to estimate the number of new overlapping intersections to process in the algorithm is valid for an average case with $m$ arbitrary clauses formed by $k$ literals of $n$ variables. Moreover, it has been also shown that the probability $P_{overlap}$ is zero for very dense clauses with many variables ($k\to n$ and $n\to \infty$). However, it is very interesting to know what happens when the density of the clauses is not 100\%, because that 
can give  an idea of the complexity of the algorithm for 
different values of $n$ variables and $k$ literals. 

Table~\ref{overlap} shows the probability $P_{overlap}$ of obtaining overlapping intersections among clauses with $k$ literals of $n$ variables and 
the number of overlapping intersections of 2 and 3 clauses (chosen from $m=n\times 100$ clauses) that require to be processed in the rest of the algorithm: ${m \choose 2} \times P_{overlap}$ and 
${m \choose 3} \times P_{overlap}$. It can be seen that with densities over 0.5, the number of new overlapping clauses added is 0 almost in all cases where $n$ is greater than 100. With densities of 0.25 it is also 0 when $n$ is greater than 800. Even with densities of 0.1 and 0.5 that number of new clauses added is 0, but in those cases $n$ has to be greater than 5000 and 20000, respectively. 

Let us think if this is an intuitive result. If we have a clause $c$ with $k$ literals of $n$ variables, and we build another one by taking variables from a box, and then deciding if the literal is positive or negative by throwing a coin, then there is a probability $\frac{k}{n}\times\frac{1}{2}=\frac{k}{2n}$ of obtaining a complementary literal for the first literal of $c$. Although there are $k$ possibilities to get a complementary one for any of the $k$ literals of $c$, the probability of overlapping is not $\frac{k^2}{2n}$ $(k\times\frac{k}{2n})$
because the extractions of variables are not independent events: the probability $\frac{k}{2n}$ of obtaining a complementary literal for the first literal changes for the following extractions that depend on the previous ones. In any case, in table~\ref{overlap} it can be noticed that when $\frac{k^2}{2n}\geq25$ (aproximately when $k\geq7\sqrt{n}$) the probability of overlapping and the number of overlapping conjunctions is 0.0000000000\footnote{The real values in table \ref{overlap} are 4.90e-25, 4.93e-15, 1.79e-14, 4.97e-13, 3.83e-12 and 7.36e-12.} (practically zero). So, the value $\frac{k^2}{2n}\geq25$ can serve as an easier way to decide if clauses are dense enough. Notice that the
$\frac{k^2}{2n}$ value for 3-SAT with 100 variables would be only 0.045, very far from 25, what means that 3-SAT clauses are not dense enough and that the
complexity of the algorithm is exponential for 3-SAT.

\subsubsection{Complexity of the algorithm with dense and disjoint clauses}
\label{complexity}

Working with such dense clauses, the algorithm processes
each clause and calculates its size. The intersections of each clause
with all the other $m-1$ clauses have to be performed, although it is only to find that they are disjoint. To process all pairs of $m$ clauses requires $O(m^2)$ time. To calculate the size of a clause requires $O(1)$ time, unless it has to be made by counting the number of literals that would require $O(k)$ time. The intersection of clauses consists on a merge of 2 sorted clauses of $k$ literals of $n$ variables that can be processed in $O(k)$ time because the literals are sorted by order of variable. Taking this into account, the complexity would be $O(m^2 \times k)$, but that is only for clauses with small values of $k$.

In fact, time complexity of the merging process is lower than $O(k)$ for clauses with bigger values of $k$. 
The merging process will stop as soon as the complementary literals are found in the sorted clauses that are being merged.  This can be approached with a geometric distribution \cite{Raikar} of a random variable that is
the number of trials (number of literals to scan in the merging process) needed to get one success (to find the first pair of disjoint literals), where the probability of success of each experiment
is $p=\frac{k}{2n}$. For the geometric distribution it is known that $1/p$ is the average or expected value, that is, the expected number of trials to find the first disjoint literal is $\frac{2n}{k}$. This means that 
for a density of 0.9 with 100 variables about 2 literals should be scanned ($\frac{2\times100}{90}=2.2$),
and for a density of 0.1 with 100 variables, 20 literals should be scanned ($\frac{2\times100}{10}=20$). Therefore, the merging process can be processed in $O(n/k)$ time\footnote{The number of literal scans does not exactly follow a geometric distribution, because $p$ is not the same for all trials. That probability $p$ grows with every fail: $\frac{k}{2n}$, $\frac{k}{2(n-1)}$, $\frac{k}{2(n-2)}$,$\ldots$ This means that the expected value is lower than $1/p=\frac{2n}{k}$. Anyway, it is also in $O(n/k)$}. 

In summary, it can be stated that SARRIGUREN is a complete algorithm for SAT that offers, in the average case, a polynomial complexity $O(m^2 \times n/k)$ for instances of $k$-SAT with $m$ clauses that are relatively dense ($k\geq7\sqrt{n}$).

	\begin{table}[h]
		\centering
		\begin{tabular}{|c|c|c||c||c||c|c|c|}
 	
			\hline   $n$ & $k$  &   $density$ & $\frac{k^2}{2n}$ & $P_{overlap}$ = $P_{ov}$ & ${m \choose 2}.P_{ov}$ & ${m \choose 3}.P_{ov}$ \\
			 &   &   $(k/n)$ & & $k_c$=$k_d$=$k$ & $k_{c,d}$=$k$ & $k_c$=$k$ \\
			 &   &  & & & &$k_d$=$1.5k$\\
\hline  10  &  9  &  0.9 & 4.0  & 0.0037109375   &  1853.6 &  2292.9 \\
\hline  100  &  90  &  0.9 & 40.5  & 0.0000000000   &  0.0 &  0.0 \\
\hline  100  &  70  &  0.7 & 24.5  & 0.0000000000   &  0.0 &  0.0 \\
\hline  100  &  50  &  0.5 & 12.5  & 0.0000001348   &  6.7 &  0.0 \\
\hline  200  &  100  &  0.5 & 25.0  & 0.0000000000   &  0.0 &  0.0 \\
\hline  200  &  50  &  0.25 & 6.2  & 0.0008437706   &  1.69e5 &  9.49e5\\
\hline  400  &  100  &  0.25 & 12.5  & 0.0000007070   &  565.6 &  5.3 \\
\hline  800  &  200  &  0.25 & 25.0  & 0.0000000000   &  0.0 &  0.0 \\
\hline  1000  &  100  &  0.1 & 5.0  & 0.0052125370   &  2.61e7 &  4.53e9 \\
\hline  5000  &  500  &  0.1 & 25.0  & 0.0000000000   &  0.5 &  0.0 \\
\hline  1000  &  50  &  0.05 & 1.2  & 0.2776349687   &  1.39e9&  1.28e13 \\
\hline  20000  &  1000  &  0.05 & 25.0  & 0.0000000000   &  14.7 &  0.0 \\
\hline
		\end{tabular} 
		\caption{Overlapping probabilities and expected number of new
		intersected clauses for different values of $k$ and $n$, where $m=n*100$}
		\label{overlap}
	\end{table} 

\subsection{Best and worst-cases for the algorithm} 

The best-cases for SARRIGUREN executed over instances of $k$-SAT with relatively dense clauses
are cases where all the clauses are disjoint among them, and the complementary literals are found during the merging as soon as possible, because those complementary literals correspond to the first variables. Some examples of best-cases are
shown in table~\ref{bestCase}. In any case, these best cases should not be much more efficient than the average cases with densities satisfying $k\geq7\sqrt{n}$, because for those cases almost all clauses should be disjoint among them. The execution of the $O(n/k)$ merging processes could be slower because the complementary literals do not necesarily correspond to the first variables.

\begin{table}[h]
	\centering
	\begin{tabular}{|c|c|c|c|c|c|c|c|}		
		\hline   $x_{1}$ & $x_{2}$ & $x_{3}$ & $\cdots$ & $x_{k-1}$ & $x_{k}$ & $\cdots$ & $x_{n}$ \\
		\hline
		\hline 0 & \multicolumn{6}{c}{\it choose $k-1$ literals with values 0 or 1} &\\		
		\hline 1 & 0  & \multicolumn{5}{c}{\it choose $k-2$ literals with values 0 or 1}&\\				
		\hline   1 & 1  & 0 & \multicolumn{4}{c}{\it choose $k-3$ literals with values 0 or 1}&\\		
		\hline  1 & 1  & 1 & \multicolumn{4}{c}{$\cdots$}&\\		
		\hline  1 & 1  & 1 & $\cdots$ & 0 & \multicolumn{2}{c}{\it choose 1 literal with value 0 or 1}&\\		
		\hline  1 & 1  & 1 & $\cdots$ & 1 & 1&\multicolumn{1}{c}{}&\\		
		\hline	
	\end{tabular} 
	\caption{Examples of best cases for the algorithm}
	\label{bestCase}
\end{table}

\begin{table}[h]
	\centering
	\begin{tabular}{|c|c|c|c|c|c|c|c|c|}		
		\hline   $x_{1}$ & $x_{2}$ & $x_{3}$ & $\cdots$ & $x_{k-1}$ & $x_{k}$ & $x_{k+1}$ & $\cdots$ & $x_{n}$ \\
		\hline
		\hline  1 & 1  & 1 & 1 & 1 & 1 &  & &\\	
		\hline  1 & 1  & 1 & 1 & 1 & & 1 & &\\	
		\hline  1 & 1  & 1 & 1 & 1 & &  & $\cdots$ &\\	
		\hline  1 & 1  & 1 & 1 & 1 & &  & & 1\\	
		
		\hline	
	\end{tabular} 
	\caption{Example of worst-case for the algorithm}
	\label{worstCase}
\end{table} 

With respect to the worst-cases, the situation is very different. That happens when all the clauses overlap among them, that is, when there are no complementary literals and the majority of the clauses have repeated literals. As the clauses are dense, that means that there will be many repeated literals in those clauses. 
In table~\ref{worstCase} it is shown an example of worst-case where
all the possible intersection of 2 clauses among the $n-k+1$ clauses overlap.
In that case the total number of possible 
intersections is in the order $O(2^{n-k+1})$, that even if clauses are 0.5 dense with $n=200$ and $k=100$ is a huge number.

At this point, it is important to say that the probability of finding $m$ random dense clauses, such as all $2^{m-1}$ posible intersections among those clauses are overlapping can be considered zero. It is obvious that if the probability of two dense clauses to overlap is 0.0000000000 (less than 1.0e-11) as shown in section \ref{clausesDoNotOverlap}, then the probability of these worst-cases for random $k$-SAT is exponentially much smaller, around $(1.0e-11)^{2^{m-1}}$.

Anyway, these pure worst-cases for SARRIGUREN working with relatively dense clauses and many repeated literals are trivial cases by using other methods. For example, in this case all the clauses are satisfied by assigning $x_1=1$. It is true that there could be other clauses with literal $\overline{x_1}$, and other clauses without that literal. In both cases, those clauses should still be dense clauses with repeated literals and they could be satisfied by using assignments for other variables. However, there could be many other dense clauses with non repeated literals, that would constitute bad cases for such assigning methods. Therefore, this should be studied further, but in any case, all these cases with overlapping clauses have a practically zero probability in random $k$-SAT with dense clauses.

\section{Experimental Results}

In this section some experimental results are presented. The algorithm has been programed in Python (see appendix \ref{algPython}), a programming language that provides support for big integers, and it has been executed in a machine with Intel Core 3.60 GHz processor, 64GB RAM, Linux Mint 19.3 Trician OS and Python 2.7. The Python program and all the input sets and obtained results in the experiments are maade available in \cite{SarrigurenSite}.

SARRIGUREN has been tested with several sets of randomly generated $k$-SAT sets of N variables (20000, 5000, 800, 200 and 100), M number of clauses (100, 1000, 10000 and 100000) and the K values corresponding to different density types (0.9N, 7RootN, 6RootN, 5RootN, 4RootN and 3RootN). The density type 0.9N is always 0.9, independently of the value of N. For example, 
the K of a 0.9N density type with N=20000 variables is 
$\lfloor 0.9\times20000\rfloor=18000$ that corresponds to a 18000/20000=0.9 density. For the rest of density types $x$RootN, the value of K is $\lfloor x\times \sqrt{N}\rfloor$. The density (K/N) depends on N, and it decreases for bigger values of N.
For example, the K for a 7RootN density type with N=5000 variables is $\lfloor 7\times \sqrt{5000}\rfloor=494$ that corresponds to a density of 494/5000=0.01. Every experiment 
for a combination of N, M and K has been repeated 20 times for almost all the cases with 100 and 1000 clauses, and for the other cases that number of executions has been reduced to 3, 2 or 1. 

\begin{figure*}
	\begin{center}
		\includegraphics[width=1\textwidth]{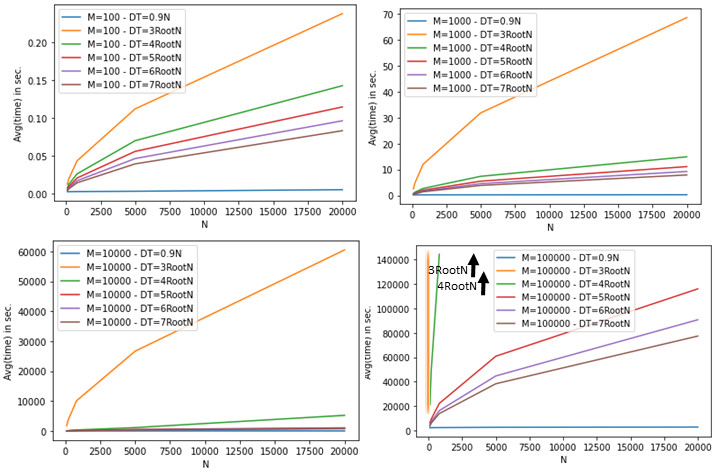} 
		\caption{Graphics for each M showing times depending on
N and density types}
		\label {graphM}
	\end{center}
\end{figure*}

\begin{figure*}
	\begin{center}
		\includegraphics[width=1\textwidth]{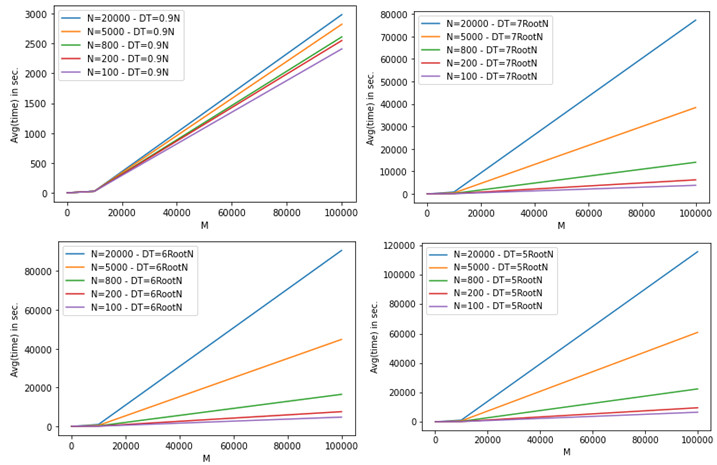} 
		\caption{Graphics for each density type showing times depending on
			N and M}
		\label{graphDT}
	\end{center}
\end{figure*}

Figures \ref{graphM} and \ref{graphDT} contain the graphics that show the average times employed by the algorithm for all the experiments, one graphic for each value of M (figure \ref{graphM}) and one graphic for the different density types (figure \ref{graphDT}). In these graphics it can be seen that the efficiency is better for higher densities: 0.9N, then 7RootN, 6RootN and going down until 3RootN. In fact for the less dense ones (4RootN and 3RootN) that do not appear in figure \ref{graphM}, some executions (with M=100000) took so long that the processes were killed before finishing. It is also easy to see that experiments were more efficient for less number of variables. Moreover, the bigger is the number of variables, then the bigger are the differences among the density types. That is because the concrete density corresponding to a density type decreases when the number of variables grows. For example, for 6RootN density type the density for 20000 variables is  
$\lfloor 6\times \sqrt{20000}\rfloor/20000=848/20000=0.042$ while the  density for 100 variables is  
$\lfloor 6\times \sqrt{100}\rfloor/100=60/100=0.6$.
Notice that for the 0.9N density type, there are no great time differences for the different number of variables, because in all cases the concrete density is the same: 0.9.

It is not surprising that the results for the 3RootN and 4RootN are very bad because the number of overlapping clauses increases a lot for those density types. In table \ref{NoOfOverlapping} it can be seen that, in average, the number of overlapping clauses for the 3RootN case grows 51 times more than the number of clauses M, and for the
4RootN case it grows more than 4 times. For the 5RootN and 6RootN cases, there are some overlapping clauses, but the efficiency does not get significatively affected. Notice that there are no overlapping clauses for the cases 7RootN and 0.9N, as explained in section \ref{clausesDoNotOverlap}.

	\begin{table}[h]
		\centering
		\begin{tabular}{|c|c|}
\hline   DT (density type) & OV\#/CLAUSES\#  \\

\hline 0.9N	& 0	   \\
\hline 7RootN &	0	   \\
\hline 6RootN &	0.0002	   \\
\hline 5RootN &	0.0664	   \\
\hline 4RootN &	4.21	   \\
\hline 3RootN &	51.4535\\
\hline

		\end{tabular} 
		\caption{Growth ratio of the number of clauses due to overlapping intersections.}
\label{NoOfOverlapping}
\end{table}

At this point it is interesting to check if the
polynomial time complexity $O(m^2 \times n/k)$ analyzed in section \ref{complexity}, is compatible with the obtained results in these experiments. That polynomial complexity was only for instances of $k$-SAT with $m$ clauses relatively dense ($k\geq7\sqrt{n}$), that is, it should be compatible at least with the experiments performed with  7RootN and 0.9N densities.

Let us start checking if the $O(m^2)$ part of the $O(m^2 \times n/k)$  complexity verifies or not. For similar values of $n/k$, when $m$ is 10 times bigger, then the expected time should be $10^2=100$ bigger. 
Table \ref{checkM2results} shows, for different M values and density types\footnote{The $n/k$ values for a density type are different, but average times of all of them are used.} DT, if the average time for an M value (1000, 10000 and 100000) is around $10^2=100$ times bigger than the average time for the M/10 value (100, 1000 and 10000). 
It can be considered compatible for the 0.9N and 7RootN density types, as expected. Moreover, it is also compatible for 6RootN and 5RootN, because the number of overlapping clauses is quite low.
 Notice that the Pearson correlation coefficient for theses cases is 0.997495124. However, it cannot be considered compatible for 4RootN and 3RootN, also as expected due to the big number of overlapping clauses.	

\begin{table}[h]
	\centering
	\begin{tabular}{|c|c|c|c||c|}

		\hline M &	DT	& Avg(time)	& Expected value:	& $O(M^2)$  \\
		&		& 	& time(M/10)*100	& compatible?\\
		\hline 100	& 0.9N &	0.00303	&&\\	
		\hline 1000	& 0.9N	& 0.24910 &	0.30381	& Yes\\
		\hline 10000 &	0.9N &	23.84	& 24.91	& Yes\\
		\hline 100000 &	0.9N &	2675.83	& 2384.17	& Yes\\
		\hline 100 &	7RootN &	0.02925		&&\\
		\hline 1000	& 7RootN &	2.84	& 2.93 & Yes\\	
		\hline 10000 &	7RootN &	275.27	& 283.55	& Yes\\
		\hline 100000 &	7RootN &	27934.50	& 27526.53	& Yes\\
		\hline 100 &	6RootN &	0.03431		&&\\
		\hline 1000	& 6RootN &	3.34	& 3.43 & Yes\\	
		\hline 10000 &	6RootN &	325.10	& 333.92	& Yes\\
		\hline 100000 &	6RootN &	32810.41	& 32509.71	& Yes\\
		\hline 100 &	5RootN &	0.04122		&&\\
		\hline 1000	& 5RootN &	4.03	& 4.12& Yes\\
		\hline 10000 &	5RootN &	389.40	& 403.50	& Yes\\
		\hline 100000 &	5RootN &	42989.82	& 38940.07	& Yes\\
		\hline 100 &	4RootN &	0.05173		&&\\
		\hline 1000	& 4RootN &	5.41	& 5.17 & Yes\\	
		\hline 10000 &	4RootN &	1398.31	& 540.85& No \\
		\hline 100 &	3RootN &	0.08506		&&\\
		\hline 1000	& 3RootN &	23.93	& 8.51	& No \\
		\hline10000	& 3RootN &	20551.57	& 2392.50	& No \\
		\hline	 
	\end{tabular} 
	\caption{Is the average time the expected one for M$^2$ and density type?}
	\label{checkM2results}
\end{table} 

Now the goal is to check if the $O(n/k)$ part of the $O(m^2 \times n/k)$  complexity is compatible or not with the experiments. For that analysis it is going to be assumed that the results of the experiments for the different configurations of N and M for the 0.9N density type
are not affected by the N/K term. In fact, they are the 
most efficient ones and its N/K value is close to 1.
Table \ref{checkNKresults} shows for the different number of variables N and density types DT (that determine the K values and therefore their N/K value) which are the average times of the experiments with all values of M and how many times  
bigger are these values compared to the values for the same configuration and density type 0.9N. 
If those results are about N/K times bigger than the corresponding result of 0.9N, then that would mean that results are compatible with the $O(n/k)$ part of the $O(m^2 \times n/k)$  
complexity. And it can be considered that it is compatible for the 7RootN, 6RootN and 5RootN density types. 
 Notice that the Pearson correlation coefficient for theses cases is 0.994232519.

Therefore, it can be affirmed that the analyzed time complexity $O(m^2 \times n/k)$ is compatible not only with the
density types 0.9N and 7RootN, but also with 6RootN and 5RootN density types. 

\begin{table}[h]
	\centering
	\begin{tabular}{|c|c|c|c|c||c|}
		
		\hline N &	DT &	Avg(time) for 	& N/K &	Ratio w.r.t. 0,9N  & $O(N/K)$ \\
		&	 &	N/K and DT	&  &case with same N & compatible?\\
		\hline 20000 &	0.9N &	752.33	& $\sim$1	&  &\\
		\hline 5000	& 0.9N	& 711.68 &	$\sim$1 &	&\\
		\hline 800	& 0.9N	& 658.89 &	$\sim$1 &	&\\
		\hline 200	& 0.9N	& 643.58 &	$\sim$1 &	&\\
		\hline 100	& 0.9N	& 608.43 &	$\sim$1 &	&\\
		\hline 20000 &	7RootN &	19520.14 &	20.22 &	25.95 & Yes\\
		\hline 5000	& 7RootN &	9691.06	& 10.12 &	13.62  & Yes\\
		\hline 800	& 7RootN &	3538.73	& 4.06 &	5.37  & Yes\\
		\hline 200	& 7RootN &	1565.04	& 2.04 &	2.43  & Yes\\
		\hline 100	& 7RootN &	950.81	& 1.43 &	1.56  & Yes\\
		\hline 20000 &	6RootN &	22886.57& 23.58 &	30.42  & Yes\\
		\hline 5000	& 6RootN &	11300.90	& 11.79 &	15.88  & Yes\\
		\hline 800	& 6RootN &	4151.87	& 4.73 &	6.30  & Yes\\
		\hline 200	& 6RootN &	1896.80	& 2.38 &	2.95  & Yes\\
		\hline 100	& 6RootN &	1187.45	& 1.67 &	1.95  & Yes\\
		\hline 20000 &	5RootN &	22886.57	& 28.29 &	38.86  & Yes\\
		\hline 5000	& 5RootN &	11300.90	& 14.16	& 21.57  & Yes\\
		\hline 800	& 5RootN &	4151.87 & 5.67	& 8.54  & Yes\\
		\hline 200	& 5RootN &	1896.80	& 2.86	& 3.72  & Yes\\
		\hline 100	& 5RootN &	1187.45	& 2.00	& 2.66  & Yes\\
		\hline		
	\end{tabular} 
	\caption{Is the average time N/K times bigger than the 0.9N case of same N?}
	\label{checkNKresults}
\end{table}

\section{Finding the solutions}
\label{solAlg}

SARRIGUREN algorithm calculates the number of solutions that satisfy a set $K$ of clauses in CNF format. The algorithm is useful to know if $K$ is unsatisfiable (when the number of solutions is 0) or satisfiable (when the number of solutions is greater than 0). It is not useful to know a concrete solution that makes $K$ satisfiable, that is, a set of literals that satisfy all the clauses in $K$. In other words, it is a complete algorithm to solve \#SAT but not an algorithm that provides the solutions to a SAT problem. But that is something that can be easily achieved by using the SARRIGUREN algorithm.

SARRIGUREN-SOL is an algorithm that gets the solutions that make $K$ satisfiable, once it is known that $K$ is satisfiable. That algorithm proceeds in this way: it selects literals corresponding to the variables in the initial set of satisfiable clauses. For each chosen literal, it checks if there are solutions for the clauses that are not satisfied with that literal. If there are solutions, then the literal is added to the solution, and if not, the complement of the literal is added to the solution. 
\begin{algorithm}[H]
	\caption{SARRIGUREN-SOL}
	\begin{algorithmic}
		\State {\bf Input:} $K = k_{1} \cdots k_{m}$, a set of disjuntive clauses
		sorted by variable number
		\State 	\hspace{1.3cm}$n$: number of variables 
		\State {\bf Output:} Set of literals that satisfy $K$
		\State {\bf Precondition:} SARRIGUREN($K$,$n$) $>$ 0, that is, $K$ is {\tt SAT}
		\State $vars \gets <x_{1},\cdots,x_{n}>$
		\State $solution \gets$ [ ]  
		\While{$vars$ is not empty {\bf and} $K$ is not empty}
		\State $lit$ $\gets$ {\it select literal ($x_i$ or $\overline{x_i}$) and remove $x_i$ from $vars$} 
		\State $K'$ $\gets$ clauses of $K$ without $lit$ or with $\overline{lit}$ which is removed
		
		\If{$\square$ is not in $K'$ {\bf and} SARRIGUREN($K'$,length($vars$)) $>$ 0} 
		\State {\bf add} $lit$ {\bf to} $solution$ 
		\Else
		\State {\bf add} $\overline{lit}$ {\bf to} $solution$ 
		\State $K'$ $\gets$ clauses of $K$ without $\overline{lit}$ or with $lit$ which is removed
		\EndIf
		\State $K$ $\gets$ $K'$
		\EndWhile
		\State {\bf return} $solution$		
	\end{algorithmic}		
\end{algorithm}

The instruction ``{\it select literal ($x_i$ or $\overline{x_i}$) and remove $x_i$ from $vars$}" has been left quite general, because different possibilities can be implemented. For example, a priority list of desired literals to be part of the solution could be provided.

Moreover, there is another algorithm that provides additional information about the solutions. The algorithm SARRIGUREN-SOL-LITS calculates the number of solutions for each literal. 

\begin{algorithm}[H]
	\caption{SARRIGUREN-SOL-LITS}
	\begin{algorithmic}
		\State {\bf Input:} $K = k_{1} \cdots k_{m}$, a set of disjuntive clauses
		sorted by variable number
		\State 	\hspace{1.3cm}$n$: number of variables 
		\State {\bf Output:} Set of literals that satisfy $K$
		\State {\bf Precondition:} SARRIGUREN($K$,$n$) $>$ 0, that is, $K$ is {\tt SAT}
		
		\State \For{{\bf each} $lit$ in $<\overline{x_{1}},x_{1},\cdots,\overline{x_{n}},x_{n}>$}
		\State $K'$ $\gets$ clauses of $K$ without $lit$ or with $\overline{lit}$ which is removed
		\State $numSolutions$ $\gets$ SARRIGUREN($K'$,$n-1$)	
		\State {\bf print} $lit$ ``has" $numSolutions$ ``solutions"
		\EndFor		
	\end{algorithmic}		
\end{algorithm}

\section{Conclusions}

In this paper a new complete algorithm for SAT (and also for SAT variations such as Unique-SAT or propositional model counting \#SAT) called SARRIGUREN has been described, analyzed and tested. The Python implementation and all the input datasets and obtained results in the experiments are made available.

That algorithm has an $O(m^2 \times n/k)$  time complexity for random $k$-SAT instances of $n$ variables and $m$ dense clauses, where dense means that $k\geq7\sqrt{n}$. With that density the number of overlapping clauses found in the algorithm can be considered zero. Notice that, under these conditions, a clause with 20000 variables and 989 literals (and therefore a density of 989/20000=0.049) is considered dense enough so that there are no overlapping clauses.
Moreover, even for less dense clauses $k\geq5\sqrt{n}$, that $O(m^2 \times n/k)$  complexity is also true because the number of overlapping clauses remains to be very small, what means that for example a clause with 20000 variables and 707 literals (0.035 density) is also considered dense. Although theoretically there could be worst-cases with exponential complexity, the probability of those cases to happen in random $k$-SAT with dense clauses can also be considered zero (even smaller than the probability of having overlapping clauses, that is zero).

One disadvantage of this complete algorithm compared to others based on inference rules is that, no understandable explanation can be offered when the set of clauses is detected as unsatisfiable; only that the $2^n$ unsatisfiable clauses have been counted.

Two complementary algorithms,
SARRIGUREN-SOL and SARRIGUREN-SOL-LITS, have also been presented that provide the solutions to $k$-SAT instances and valuable
information about number of solutions for each literal.

Although SARRIGUREN does not solve the NP=P problem because it is not a polynomial algorithm for 3-SAT, it broads the knowledge about that subject. The assumption that $k$-SAT does not have subexponential algorithms for $k \geq 3$ has to be revisited, according to this work.

Finally, it has not escaped my knowledge that SARRIGUREN can be modified in order to obtain new heuristic SAT algorithms and parallel SAT algorithms.

On the one hand, for the case of heuristic SAT algorithms, it is quite obvious that if $k\geq5\sqrt{n}$, the probability of finding overlapping clauses is very close to zero (practically zero if $k\geq7\sqrt{n}$). In that case, the part of the algorithm where intersections are calculated can be omitted, and the complexity 
would drastically drop to a linear $O(m)$. SARRIGUREN-SOL algorithm would provide solutions that could be easily tested, also with $O(m \times n)$ complexity. 

On the other hand, for the case of parallel algorithms, it is easy to see that
the number of clauses can be calculated by levels.
In a first level, $\sum_{i=1}^{m} |\{\overline{k_{i}}\}|$ can be calculated by processors that compute $|\{\overline{k_{i}}\}|$  in parallel and report the results to a node that performs the sum. Once that level is finished, the same can be done to compute the negative cardinalities $-\sum_{j,k: 1\leq j<k\leq m} |\{\overline{k_{j}}\}\cap \{\overline{k_{k}}\}|$. And after that, the level that calculates again positive cardinalities: $+\sum_{j,k,l: 1\leq j<k<l\leq m} |\{\overline{k_{j}}\}\cap \{\overline{k_{k}}\} \cap \{\overline{k_{l}}\}|$, and so on. Notice that, if after the completion of a level that reports negative cardinalities the value $2^n$ is reached, then the parallel computation can end and return UNSAT.

\appendix
\section{Python code of SARRIGUREN}
\label{algPython}
{\scriptsize
\begin{verbatim}
def pattern(c):                   def sarriguren(K,n,getNumOverlapping=False):                
   return c                       # Precondition: K disjunctive clauses  
                                  #               sorted by variable number
def cardinality(c,n):                P=[]     
   return 2**(n-len(c))              u=0
                                     m=len(K) 
def contrary(sign):                  for i in range(0,m):              
   return '+' if sign=='-' else '-'     ap=pattern(K[i])  
                                        N=[]  
def intersect(c,d):                     u+=cardinality(ap,n)
   result = []                          N.append(['-',ap])
   i = j = 0                            for sp in P: 
   while i < len(c) and j < len(d):        ip=intersect(ap,sp[1]) 
      if c[i] == -d[j]:                       if ip!=[]:
         return []                               if sp[0]=='+': u+=cardinality(ip,n)
      elif c[i]==d[j]:                           else: u-=cardinality(ip,n)
         result.append(c[i])                     N.append([contrary(sp[0]),ip])
         i += 1                         P=P+N     
         j += 1                         if u==2**n:         
      elif abs(c[i])<abs(d[j]):            if getNumOverlapping:       
         result.append(c[i])                  return [0,len(P)-m]   
         i += 1                            return 0
      else:                          if getNumOverlapping:          
         result.append(d[j])            return [2**n-u,len(P)-m]         
         j += 1                      return 2**n-u                
   result.extend(c[i:])                            
   result.extend(d[j:])                      
   return result                             
                                             def sarriguren_sol(K,n,vars=[]): 
def filterClauses(K,l):                      # Precondition: sarriguren(K,n)>0  
   Kprime=[]                                 # vars is priority list of literals to assign 
   for c in K:                               # (it must have exactly the n literals)
      if -l in c:                               solution=[]
         d=[]                                   if vars==[]: # random literals tried
         for lit in c:                            for in in range(1,n+1):
            if lit!=-l: d.append(lit)                if random.randint(0,1)==0: vars.append(i)
         if d==[]: return False # K UNSAT            else: vars.append(-i)
         Kprime.append(d)                       while vars!=[] and K!=[]:      
      elif l not in c:                            l=vars[0] 
         Kprime.append(c)                         vars.remove(l)
   return Kprime                                  Kprime=filterClauses(K,l) 
                                                  if Kprime and sarriguren(Kprime,len(vars))>0: 
def sarriguren_sol_lits(K,n):                        solution.append(l) 
# Precondition: sarriguren(K,n)>0                 else: 
   Sols=[]                                           solution.append(-l)       
   for lit in range(1,n+1):                          Kprime=filterClauses(K,-l)                    
      for sign in [-1,1]:                         K=Kprime                        
        Kprime=filterClauses(K,lit*sign)        return solution        
        if Kprime==False: numSolutions=0  
        else: numSolutions=sarriguren(Kprime,n-1)                                 
        Sols.append([lit*sign,numSolutions])                                   
  return Sols                                 

\end{verbatim}
}
\bibliography{sarriguren}
\bibliographystyle{acm}

\end{document}